\documentclass[10pt,aps,twocolumn,longbibliography,superscriptaddress,floatfix]{revtex4-2}

\usepackage{graphicx} 
\usepackage{bm}
\usepackage{color}
\usepackage[table]{xcolor}
\usepackage{makecell}
\usepackage{subfigure}
\usepackage{amsmath}
\newcommand{\xdownarrow}[1]{%
  {\left\downarrow\vbox to #1{}\right.\kern-\nulldelimiterspace}
}
\usepackage{contour}
\usepackage[normalem]{ulem}
\usepackage{comment}
\usepackage{amssymb}
\usepackage[percent]{overpic}
\usepackage{comment}
\usepackage{bbold}
\usepackage{enumerate}
\usepackage{float}
\usepackage{xspace}
\usepackage{mathtools}
\usepackage{mathrsfs}
\usepackage{mathptmx}
\usepackage[colorlinks=true,linkcolor=blue,citecolor=blue,urlcolor=blue]{hyperref}
\usepackage{orcidlink}
\usepackage{tikz}
\usepackage{siunitx}
\usepackage[utf8]{inputenc}
\usepackage{nicefrac}
\usepackage[capitalise]{cleveref}
\Crefformat{figure}{#2Fig.~#1#3}
\Crefformat{equation}{#2Eq.~(#1#3)}

\newcommand{\smsz}{\ensuremath{\text{S}^z}}
\newcommand{\smJij}{\ensuremath{J_{ij}}\xspace}
\newcommand{\smmu}{\ensuremath{\mu_{\mathrm{s}}}\xspace}

\newcommand{\smB}{\ensuremath{\mathbf{B}}\xspace}

\newcommand{\smA}{\ensuremath{A}\xspace}

\newcommand{\sms}{\ensuremath{\mathbf{S}}\xspace}
\newcommand{\vampire}{\textsc{vampire} }

\newcommand{\muB}{\ensuremath{\mu_{\mathrm{B}}}\xspace}

\begin{document}
\title{Intrinsic Spin Nernst Effect and Chiral Edge Modes in van der Waals Ferromagnetic Insulators: Dzyaloshinskii-Moriya {\it{vs.}} Kitaev Interactions}

\author{Verena Brehm \orcidlink{0000-0002-7174-1899}}
\email{verena.j.brehm@ntnu.no}
\affiliation{Center for Quantum Spintronics, Norwegian University of Science and Technology, 7034 Trondheim, Norway}
\author{Pawe\l{} Sobieszczyk\orcidlink{0000-0002-1593-1479}} 
\affiliation{Institute of Nuclear Physics Polish Academy of Sciences, Radzikowskiego 152, 31-342 Krakow, Poland}
\author{Alireza Qaiumzadeh \orcidlink{https://orcid.org/0000-0003-2412-0296}} 
\affiliation{Center for Quantum Spintronics, Norwegian University of Science and Technology, 7034 Trondheim, Norway}

\begin{abstract}
The thermomagnetic Nernst effect and chiral edge states are key signatures of nontrivial topology and emerging Berry curvature in magnonic systems. Implementing atomistic spin simulations, we theoretically demonstrate the emergence of chiral magnon edge states at the boundaries of a ferromagnetic hexagonal lattice in the presence of Dzyaloshinskii-Moriya and Kitaev interactions, which are robust against nonlinear magnon interactions. In our simulations, we consider the spin parameters of CrI$_3$ as a prototype of van der Waals magnetic layers. We show that the spin accumulation is reduced in the presence of Kitaev spin interactions compared to systems governed by Dzyaloshinskii-Moriya interactions. This reduction stems from the breaking of the $U(1)$ symmetry, which leads to a shorter spin coherence length imposed by the Kitaev interaction. We propose that measuring the angular dependence of the Nernst signal in a magnetic field provides an effective indirect method for identifying the microscopic origin of topological magnons. 
Our findings hold promising potential for advancing next-generation energy-harvesting Nernst materials and facilitating the integration of topological magnetic materials with spintronic-based quantum technologies.

\end{abstract}

\maketitle

\section{Introduction}
The large family of Hall effects encompasses a rich variety of exotic phenomena in both fermionic and bosonic condensed matter systems. These effects not only reveal essential insights into the topological nature of these systems but also serve as probes for exploring the quantum geometry tensor, anomalies in quantum field theory, the nature of disorders, topological magnetic textures, and magneto-optic properties of the system 
\cite{NiuBerryPhase,NagaosaAnomalousHallEffect,SpinHallEffects,Du2021nonlinHalleff,CHERNODUB20221,TopQuantumHallEff,HatamiDisorder,Varshney2023,PhysRevB.107.014410,PhysRevB.110.054431,ovalle2024orbitalkerreffectterahertz}.
Among the various Hall effects, the thermo-electro-magnetic Hall effects, commonly known as the Nernst effect, are notable for their ability to convert a temperature gradient, generated by sources such as waste heat, electric current, or heat pulses, into transverse heat, charge, orbital, and spin signals. This phenomenon has great potential, not only for technological applications, but also for advancing our understanding of the fundamental interactions in quantum condensed matter systems \cite{BehniaNernstEff,mizuguchiEnergyHarvestingMatNernst,CHERNODUB20221,FelserReviewTopMat,spinNernstQuantumEntanglement,PhysRevB.107.014410}.

The discovery of the magnon Hall effect in a three-dimensional (3D) magnetic insulator in 2010 marked a significant breakthrough, opening a new avenue for exploring the Nernst effect by replacing conventional charge carriers with low-loss magnons \cite{OnoseMagnonHallEffScience}.
Magnetic materials can host topological magnons \cite{McClarty_2022}, whose influence on quantum transport manifests through phenomena such as the thermal magnon Hall effect and the magnon spin Nernst effect, as well as the emergence of symmetry-protected topological chiral edge and hinge states \cite{TheOriginalLuttinger,OjiThermTransportCoeff,OnoseMagnonHallEffScience,MatsumotoThermalHallEffect,mookThermHallEff1,MookMagHallEffKagome,HirschbergerThermHallEffKagome,OwerreTopHoneycombMagHallEff,KwonHaldaneKaneMeleModel,ExpSNE_1_Meyer2017,ExpSNE_2_Sheng2017,ExpSNE_3_Bose2018,joshi,McClartyThermHallEffKitaev,Kovalev,Thermal_Hall_Effect_Spin_Nernst_Effect_Spin_Density,RuClthermalHallEff,weißenhofer2024atomisticspindynamicssimulations,ExpThermHallCrI3,MagnonOrbital,Toplogical_chiral_magnon_edge_mode_in_magnonic_crystal}

Recently discovered two-dimensional (2D) van der Waals (vdW) magnets \cite{HuangNature2017,Mak2DlayeredMagMat, GibertiniMagnetic2Dmaterials}, known for their highly tunable magnetic and electronic properties \cite{AlirezaDFT,  Ebrahimian_2023,PhysRevMaterials.8.054002,zhong2024integrating2dmagnetsquantum}, offer a promising platform for investigating chiral magnons, topological magnon states, and magnon Hall effects in 2D systems. 
The semiconducting ferromagnet CrI$_3$ layer with a honeycomb magnetic lattice is a prototype of these materials \cite{RamanFePs3Wang2016,Gong2017intrinsicFMorderInvdWcrystal,HuangNature2017,Biquadratic_exchange_interactions_in_2D_magnets,Elton_QuantumRescaling2021,Genome22,Hicken22,Jenkins22,PhysRevB.106.054403,Coronado23,Yonathan23,Hicken23,Srini23,TANG20231}.
CrI$_3$ has attracted attention after the experimental observation of a band gap in the magnon dispersion relation at the $K$ symmetry points \cite{ChenPRX2018,ChenPRX2021}. The microscopic origin of this gap is subject to debate. Although a nontopological origin could lie in electron correlations and spin-phonon interactions in multilayer structures \cite{Ke2021,Magnon_Phonon_Delugas2023}, many theoretical studies propose a topological origin in their 2D single layers from either Dzyaloshinskii-Moriya (DM) \cite{ChenPRX2018,Biquadratic_exchange_interactions_in_2D_magnets,ChangsongDFT,ChenPRX2021,milosevicPRB,Theory_of_magnetism_in_the_vdW_magnet_CrI3,Soenen_Multimodel2023} or Kitaev \cite{Xu2018Anisotropy,Hammel_FundamentalSpinInterac2020,Theory_of_magnetism_in_the_vdW_magnet_CrI3,HPTkitaev2020Aguilera,PerturbTheoryKitaevStavropolos2021,Elliot2021,Soenen_Multimodel2023,ChenKee2023_Kitaev,robertoMagPhononTop} interactions. Both of these interactions lead to topologically nontrivial magnon bands and thus to the thermal Hall effect of magnons and the magnon spin Nernst effects \cite{Kovalev}. 

Although Hall effects in fermionic systems can be tuned by changing the Fermi level to cross the topological gap and can be experimentally explored by measuring electric voltage, their bosonic counterparts present significant challenges.
In this article, we investigate chiral magnon edge states and the spin accumulation that arises from the magnon Nernst effect in a confined geometry as an observable and direct probe of the topological properties of 2D vdW magnets. We compare two spin models, the DM and Kitaev models, which generate a topological bulk magnon gap. Solving the stochastic Landau-Lifshitz-Gilbert (sLLG) equation, we incorporate nonlinear magnon interactions to assess the robustness of edge modes and spin accumulations in the presence of such nonlinear effects. We investigate the spin accumulation on chiral topological edge states in both out-of-plane (OOP) and in-plane (IP) magnetic states, set by an external magnetic field. Tuning the magnetization direction can alter the topological properties of the DM and Kitaev models in different ways \cite{DMIandKitaevinCrI3analytical,PhysRevB.105.L100402,kløgetvedt2023tunable,soenenTunableMagTop}. 
By comparing the transverse spin signals of these two spin models in response to a longitudinal temperature gradient, we present an experimentally accessible direct probe of the underlying microscopic interactions and topological characteristics in vdW magnetic materials. 

Chiral edge modes can be excited directly using a monochromatic magnetic field pulse, with fine tuning of the field frequency \cite{MookHingeModes}. However, for practical applications, it is essential to rely on quantum transport techniques, as they provide better control and manipulation of these modes while remaining compatible with modern nanotechnologies. Applying a temperature gradient, we excite both the bulk and edge modes in a ferromagnetic layer with confined geometry. Without loss of generality, we use the spin parameters of CrI$_3$ as the prototype of 2D vdW ferromagnets. The bulk-boundary correspondence dictates that bulk properties, such as a finite intrinsic Nernst response, are linked to the presence of edge states. As a result, we measure the corresponding spin accumulation at the boundaries, which provides direct insight into the interplay between bulk phenomena and edge excitations. Very recently, topological magnon edge states in CrI$_3$ were experimentally observed using scanning tunneling microscopy \cite{zhang2024directobservationtopologicalmagnon}.

\section{Methods}
\subsection{Model}
Building on our previous study \cite{PRBcri3article}, we compare the Kitaev spin model with the DM spin model to describe the spin dynamics in CrI$_3$. 
For the DM spin model, we consider the following spin Hamiltonian \cite{Biquadratic_exchange_interactions_in_2D_magnets},
\begin{equation}
\begin{aligned}\label{eq:H_DM}
\mathscr{H}_\text{DM} = &- \sum_{i,j} \big(\smJij \sms_i \cdot \sms_j + \lambda_{ij} \smsz_i \smsz_j \big) - D_z \sum_i \left(\smsz_i\right)^2  \\& - \smmu h_0 \sum_i \smB \cdot \sms_i   - K_\text{bq}\sum_{\langle i,j\rangle}\left(\sms_i \cdot \sms_j \right)^2 \\ &- \smA_z\sum_{\langle\langle i,j\rangle\rangle} \nu_{ij}  (\sms_i \times \sms_j)^z,
\end{aligned}
\end{equation}
while for the Kitaev spin model, we use the following spin Hamiltonian  \cite{Hammel_FundamentalSpinInterac2020},
\begin{equation}
    \begin{aligned} \label{eq:H_Kitaev}
    \mathcal{H}_{\kappa} = &- J_0\sum_{\langle i,j\rangle} \sms_i \cdot \sms_j  - D_z \sum_i \left(\smsz_i\right)^2 - \smmu h_0 \sum_i \smB \cdot \sms_i  \\&- \kappa \sum_{\langle i,j\rangle \in \eta } \text{S}_i^\eta \text{S}_j^\eta .
\end{aligned}
\end{equation}
Both models are capable of describing the magnon dispersion of CrI$_3$ with an OOP magnetization direction. Additionally, a similar Kitaev model was recently applied to fit the excitation spectra of low-energy magnons in the vdW ferromagnet VI$_3$ \cite{PhysRevLett.132.246702}.
In the Hamiltonians~(\ref{eq:H_DM}) and (\ref{eq:H_Kitaev}), the direction of the magnetic moment at site $i$ is carried by the unit vector $\sms_i$  \cite{vampire}, and the direction of the applied magnetic field with strength $h_0$ is denoted by \smB. The magnetic field is applied along either the OOP ($z$) direction or the IP ($x$) direction. $\mu_s$ is the atomic magnetic moment. The magnetic ground state in the absence of a magnetic field is OOP along the $z$ direction, determined by the easy-axis magnetic anisotropy with strength $D_z>0$ . $\langle ... \rangle$ and $\langle \langle ... \rangle \rangle$ denote sums over nearest-neighbor (NN) and next-nearest-neighbor (NNN). 

\begin{figure}
    \centering
    \includegraphics[trim=0 50 0 50,clip,width=\linewidth]{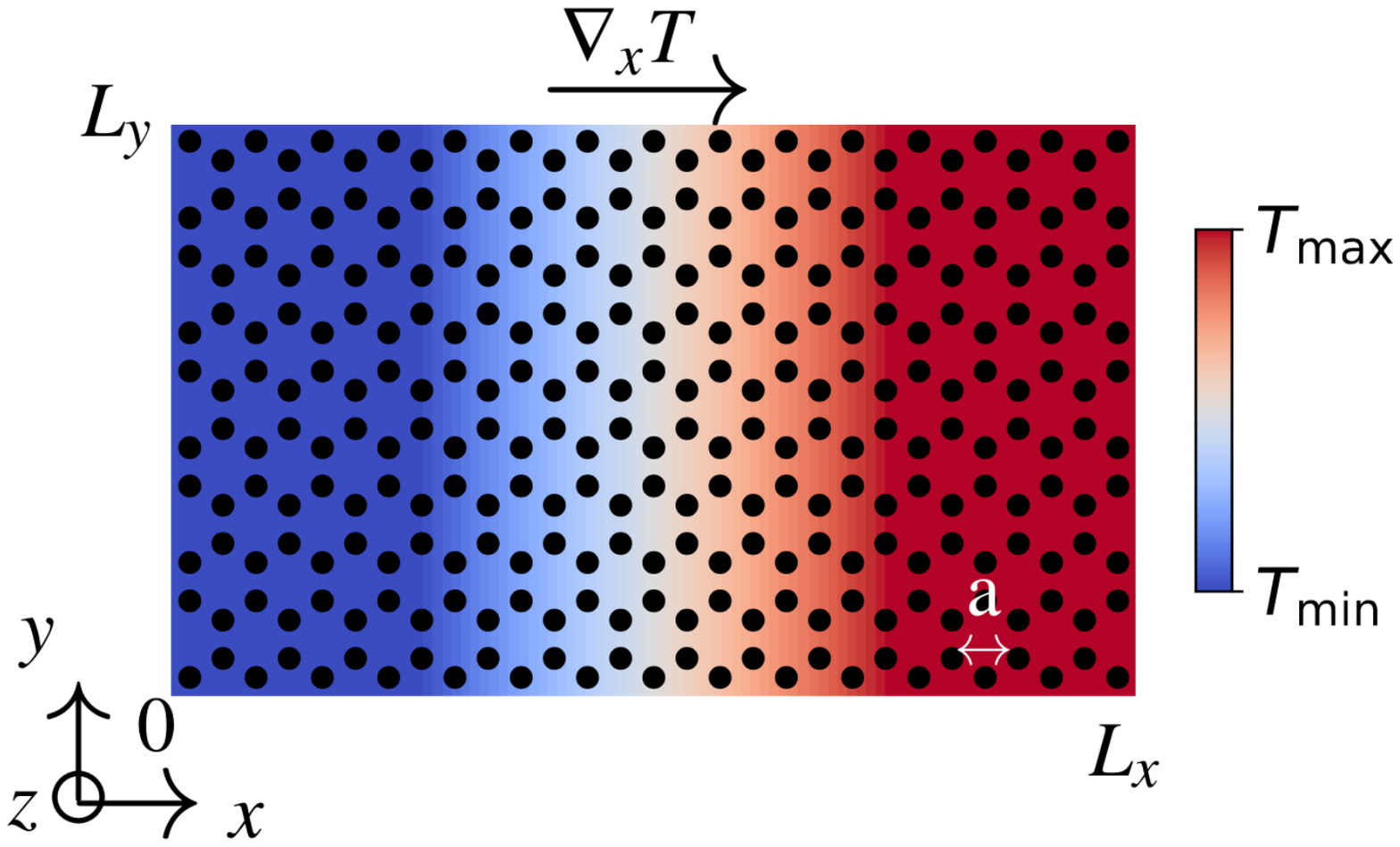}
    \caption{Schematic illustration of the system setup: We study a FM single layer with a honeycomb lattice structure, characterized by zigzag edges at the top and bottom boundaries ($x=0$ and $x=L_x$) and armchair edges on the left and right boundaries ($y=0$ and $y=L_y$). A temperature gradient is applied along the $x$ direction. The red-blue color bar represents the thermal gradient across the layer. 'a' denotes the lattice constant.}
    \label{fig:lattice}
\end{figure}

In the DM spin model (\ref{eq:H_DM}), the NNN OOP DM interaction, with an amplitude $\smA_z$ and the Haldane sign $\nu=+(-)1$ for clockwise (counterclockwise) hopping \cite{Soenen_Multimodel2023,milosevicPRB}, opens a topologically nontrivial magnon band gap between the acoustic-like and optical-like magnon branches at the $K$ symmetry points, as long as the magnetization direction is not orthogonal to the DM vector \cite{PRBcri3article}. $J_{ij}$ and $\lambda_{ij}$ parameterize isotropic and anisotropic bilinear Heisenberg exchange interactions up to the third NNs, while $K_{\rm{bq}}$ refers to the biquadratic exchange interaction.

In the Kitaev spin model (\ref{eq:H_Kitaev}), on the other hand, the NN bond-$\eta$-dependent Kitaev interaction with strength $\kappa$ \cite{McClartyThermHallEffKitaev} opens a topologically nontrivial band gap in the magnon dispersion \cite{PRBcri3article}. The Kitaev term is added to the NN isotropic Heisenberg exchange interaction with the strength $J_{0}$ in this spin Hamiltonian.

It is worth mentioning that even though both the DM and Kitaev interactions lead to the same topological magnon gap at $K$ symmetry points with the same topological Chern number, they demonstrate some subtle differences.
First, in the interaction of DM with the OOP DM vector, the $x-y$ plane is isotropic and preserves the $U(1)$ symmetry. In this model, the topological gap is closed if the magnetization lies inside this plane. However, the Kitaev interaction introduces an anisotropy in the $x-y$ plane and thus breaks the $U(1)$ symmetry. In this model, the closing of the topological gap occurs at different polar angles, which depend on the IP azimuthal angle of the magnetization direction \cite{PRBcri3article}.
Second, while the DM spin model, in the absence of a magnetic field, commutes with the \( z \)-component of the spin angular momentum, the Kitaev spin interaction does not. Consequently, the \( z \)-component of the spin angular momentum is not a good quantum number in the Kitaev model. The implications of these distinctions will be evident in the transport results discussed in the following.

\subsection{Magnon spin Nernst effect and edge spin accumulation} 
A temperature gradient induces a transverse magnon current, leading to spin accumulation at the system boundaries if the magnon spin Nernst coefficient, $\alpha_{xy}$, is nonzero. This coefficient relates the transverse magnon current density, ${j}_y^m$, to the longitudinal temperature gradient, ${\nabla}_x T$, through the relation ${j}_y^m = \alpha_{xy} {\nabla}_x T$ \cite{Thermal_Hall_Effect_Spin_Nernst_Effect_Spin_Density,KovalevNernstEffDMIfm}. We investigate the \emph{intrinsic} magnon spin Nernst effect, which arises from a finite \emph{spin} Berry curvature in the magnon bands.
Both Kitaev and DM interactions can lead to the emergence of a finite spin Berry curvature and thus a nontrivial magnon topology \cite{DMIandKitaevinCrI3analytical}. 
However, both DM and Kitaev interactions not only lead to the same topological gap, but also result in similar Berry curvature for the OOP configuration. The analytical formula for finding $\alpha_{xy}$ describing the intrinsic Nernst effect in the bulk depends only on the Berry curvature and Bose-Einstein distribution and thus cannot capture the $U(1)$ breaking of the Kitaev interaction which leads to a short magnon coherency. Thus, our numerical approach gives insight into realistic transport results beyond analytical scope.

The transverse magnon current induces a spin accumulation at the edges. This spin accumulation can be measured, from the spin pumping relation \cite{firstSpinPumping}, with
    \begin{equation} \label{eq:spinAccum}
        \bm{\mu}_i :=  g_r^{\uparrow \downarrow}  \left<\bm{S}_{i}(t)\times \dot{\bm{S}}_{i}(t)\right>_t,
    \end{equation}
where $\bm{\mu}_i$ is the vector of spin accumulation at site $i$, $g_r^{\uparrow \downarrow}$ is the real part of the spin mixing conductance \cite{ArneSpinMixingConductance}, and $\left<\cdot \right>_t$ denotes a time average. We are interested in the dc spin accumulation, which is represented by the component of $\bm{\mu}$ along the magnetic ground state, i.e., the $z$ component, $\mu^z$, when the magnetization is OOP and the $x$ component, $\mu^x$, when the magnetization is IP. In our calculations, we measure $\mu$ in units of $g_r^{\uparrow \downarrow}$. 
Scattering of nonequilibrium magnons at the interface leads to the spin accumulation at the boundaries, described by
${\bm{\mu}}_i$. In the linear response regime this quantity is proportional to the average deviation of the magnetic moment $\langle \delta \bm{S}_i\rangle$. While the former can be electrically measured via the inverse spin Hall effect in an adjacent heavy metal layer, the latter can be directly read out optically.

\subsection{Atomistic spin model simulations}
To simulate CrI$_3$ magnetic layers, we set up a 2D honeycomb lattice of finite size $L_x=L_y=\SI{100}{nm}$ with nonperiodic boundary conditions, presented schematically in \cref{fig:lattice}. The edges are zigzag along the $x$ direction (top and bottom boundaries) and the armchair along the $y$ direction (left and right boundaries).
In the main text, we focus on spin accumulation at the zigzag edges, where the temperature gradient is applied along the $x$ direction. The results for the armchair edges, where the temperature gradient is along the $y$ direction, are presented in \cref{sec:armchairEdges}. Although the dispersion of the edge modes is different for both models on the zigzag edge compared to the armchair edge \cite{MookLossQuBitcoupling}, for the DM model, there is no significant difference in the transverse spin signal. However, in the Kitaev model, the transverse spin signal is quite different on the armchair edges and we observe indications of possible magnon corner states, a signature of higher-order topological states, which require further investigation and are beyond the scope of the present article.

The spin dynamics in both the DM and Kitaev models are separately simulated  using atomistic spin simulations by solving the nonlinear sLLG equation, implemented via the \vampire software package. 
For technical details, see \cref{sec:technicalDetails}.

\begin{figure}
    \centering
    \includegraphics[width=\linewidth]{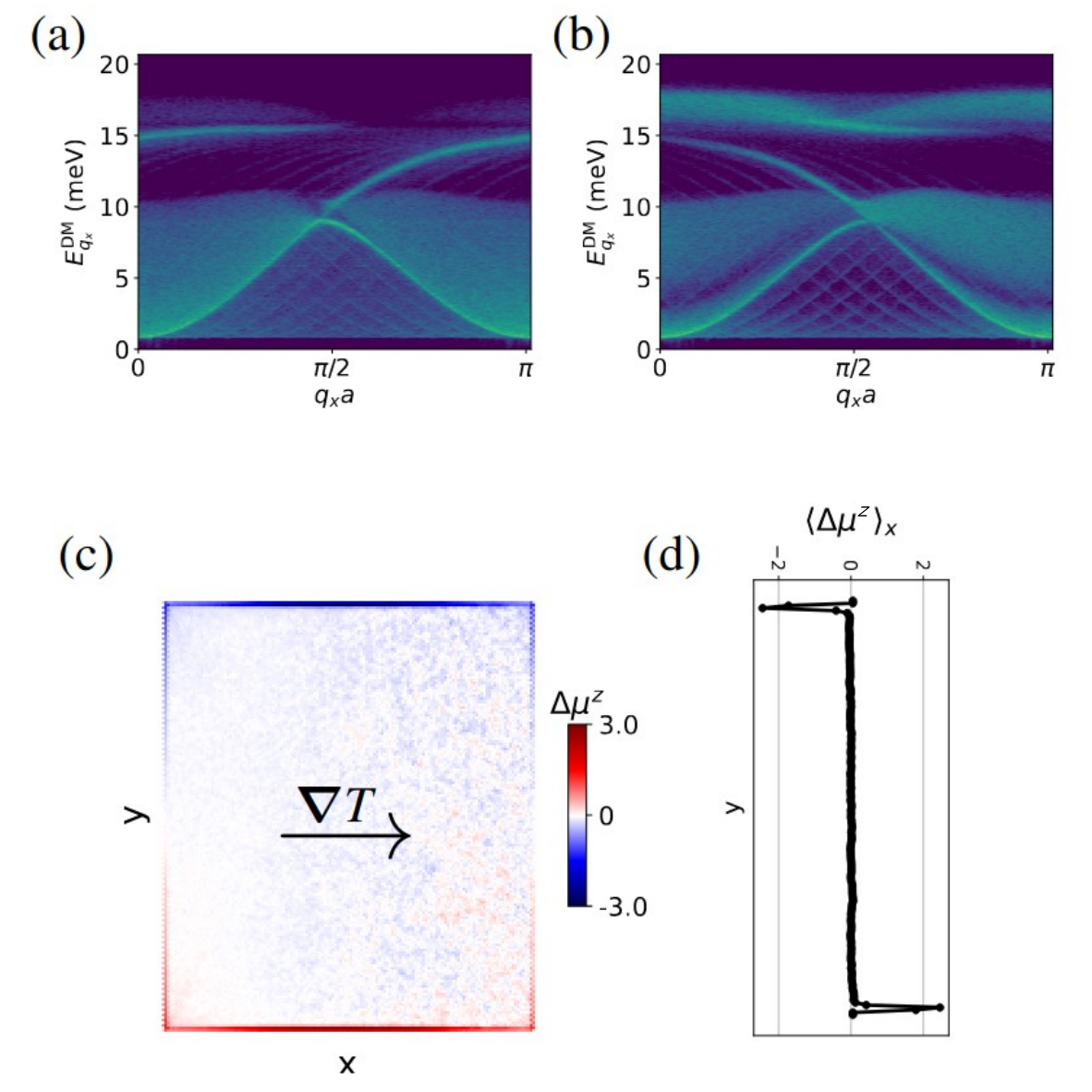}
    \caption{Response of the spin system to a temperature gradient along the $x$ direction in the DM model with OOP magnetization. (a) and (b) show the bulk magnon dispersion and the chiral edge modes of the upper zigzag edge (a) and lower zigzag edge (b) inside the topological magnon gap at the $K$ symmetry point. (c) shows the 2D spatial distribution of the $z$ component of nonequilibrium spin accumulation. The red-blue color bar indicates the spin accumulation $\Delta \mu>0$, and depletion $\Delta \mu<0$. (d) illustrates the 1D spatial distribution of the spin accumulation as a function of the transverse $y$ direction, obtained by averaging the spin accumulation over the longitudinal 
 $x$ direction.}
    \label{fig:DM_OOP}
\end{figure}

\section{Results}
\subsection{DM spin model} 
First, we demonstrate how the spin system, described by the DM model, Eq.~(\ref{eq:H_DM}, responds to the application of a temperature gradient.
As mentioned above, the DM interaction introduces a topological magnon gap in the bulk magnon dispersion. In our finite-size geometry, chiral edge states appear inside the topological magnon gap as shown in Figs. \ref{fig:DM_OOP}(a) and (b). We present the magnon dispersion at the top \ref{fig:DM_OOP}(a) and bottom \ref{fig:DM_OOP}(b) zigzag edges, when the magnetization is OOP, in the presence of the applied temperature gradient. The color map in the magnon dispersion plots represents the intensity of the thermal magnon density distribution. We see that both the bulk and edge magnon modes are thermally populated in our system. 
A change in the slope of the edge modes at opposite edges is a signature of their chiral nature.

The spatially resolved distribution of the nonequilibrium magnon accumulation for the OOP magnetic configuration is shown in \cref{fig:DM_OOP}(c). 
In all spin accumulation plots, we subtract the signal of the thermal equilibrium background from the total signal to measure only the nonequilibrium contribution $\Delta \mu$, in units of \SI{}{\per\femto\second}, see \cref{sec:technicalDetails} for more details. On average, the deviation of the spin accumulation at the edges corresponds to 10\% of the bulk value.

The clear difference in the spin accumulation at opposite edges indicates the chirality of the edge states. In \cref{fig:DM_OOP}(d) a one-dimensional (1D) profile of the spin accumulation is shown, found from an average along the longitudinal $x$ direction. A large spin signal can be seen at the edges with opposite signs at the two edges. Note that in ferromagnets, there is only one type of polarized magnons, and $\Delta \mu > (<) 0$  denotes spin accumulation (depletion).
Although both boundaries support edge modes with opposite chiralities, applying a temperature gradient in a specific direction leads to an unequal population of the chiral edge modes, favoring one over the other.

By applying a strong magnetic field along the $x$ direction, the magnetization is tilted IP. Since the DM vector is OOP, this means that the topological magnon gap at the $K$ symmetry points is closed \cite{PRBcri3article}. Hence, the system is topologically trivial in this case, and neither edge modes nor transverse transport exist; see \cref{sec:DMwithIPmag}. It has been proposed that in CrI$_3$, grown on some substrates, the DM vector might be tilted. As a result, tilting the magnetization IP does not close the topological gap \cite{PRBcri3article}.

\subsection{Kitaev spin model}
\begin{figure}
    \includegraphics[width=\linewidth]{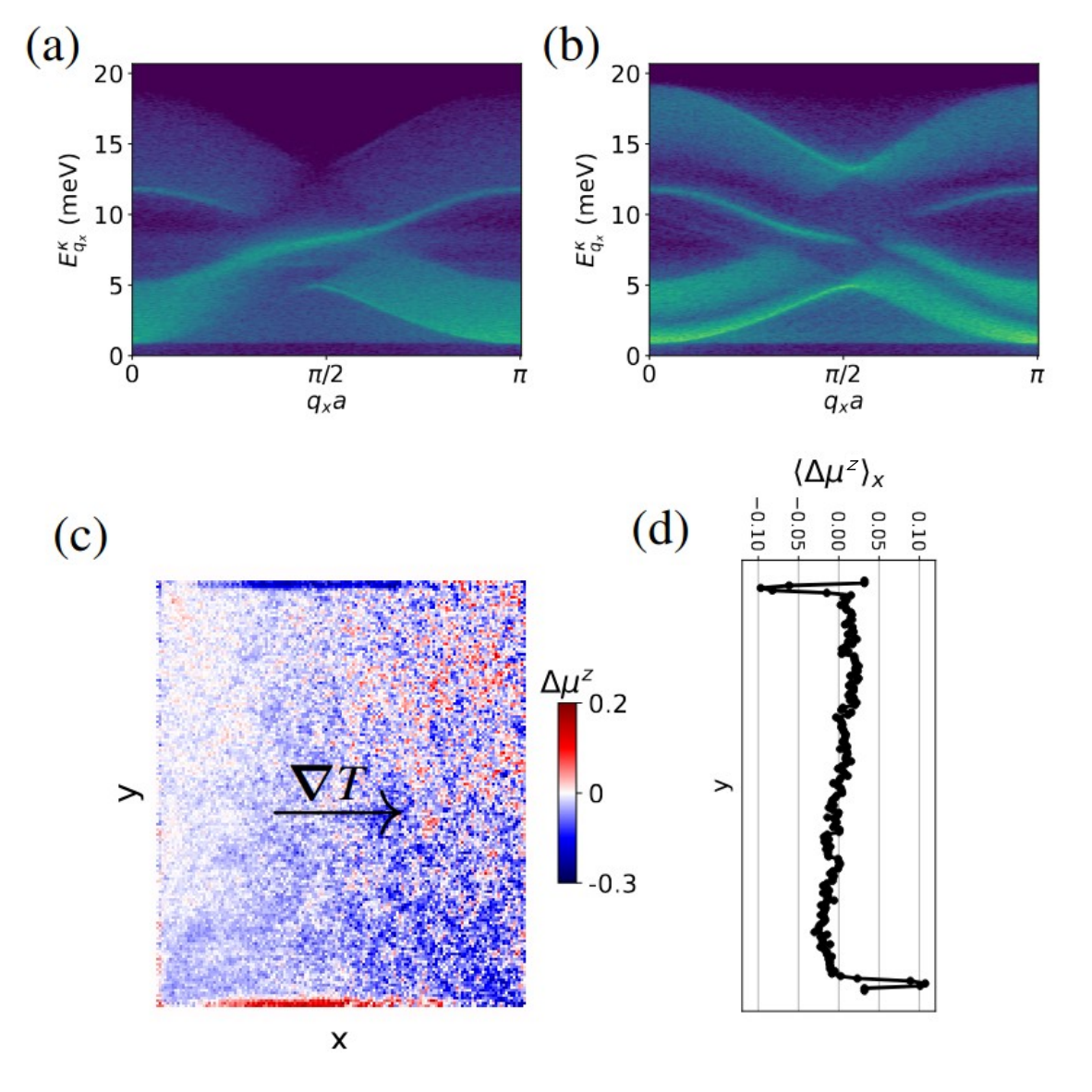}
    \caption{Response of the spin system to a temperature gradient along the $x$ direction in the Kitaev model with OOP magnetization. (a) and (b) show the bulk magnon dispersion and the chiral edge
modes of the upper zigzag edge (a) and lower zigzag edge (b) in-
side the topological magnon gap at the $K$ symmetry point. (c) presents the 2D spatial distribution of the $z$ component of nonequilibrium spin accumulation. (d) shows the spin accumulation as a function of the transverse $y$ direction, averaged along the longitudinal $x$ direction.}
    \label{fig:K_OOP}
\end{figure}

Now, we shift our focus to the Kitaev spin model, Eq. (\ref{eq:H_Kitaev}), with OOP magnetization \cite{joshi}.
The bulk magnon gap at the $K$ symmetry points and the chiral edge states, appearing inside the topological gap, are shown in \cref{fig:K_OOP}(a) and (b) in the presence of the applied temperature gradient. Similar to the DM spin model with OOP magnetization, the change in slope of the edge mode inside the bulk gap on the upper and lower edges demonstrates the chirality of the edge modes. 

In \cref{fig:K_OOP}(c) and (d), nonequilibrium spin accumulations with opposite signs can be seen at the two transverse edges. However, the signal is much weaker compared to the DM model, see \cref{fig:DM_OOP}, although we have used the same Gilbert damping parameter and the same normalized temperature gradient (for technical details, see \cref{sec:technicalDetails}). 
This effect can be attributed to the nature of the Kitaev spin interaction that breaks $U(1)$ symmetry and does not conserve the spin angular momentum. This leads to a significant reduction in the spin coherence length of magnons. Consequently, spin accumulation at the edges is greatly suppressed.
We conclude that while a chiral edge signal can exist in the Kitaev model, it is significantly suppressed compared to the DM model.

\begin{figure}
    \includegraphics[width=0.95\linewidth]{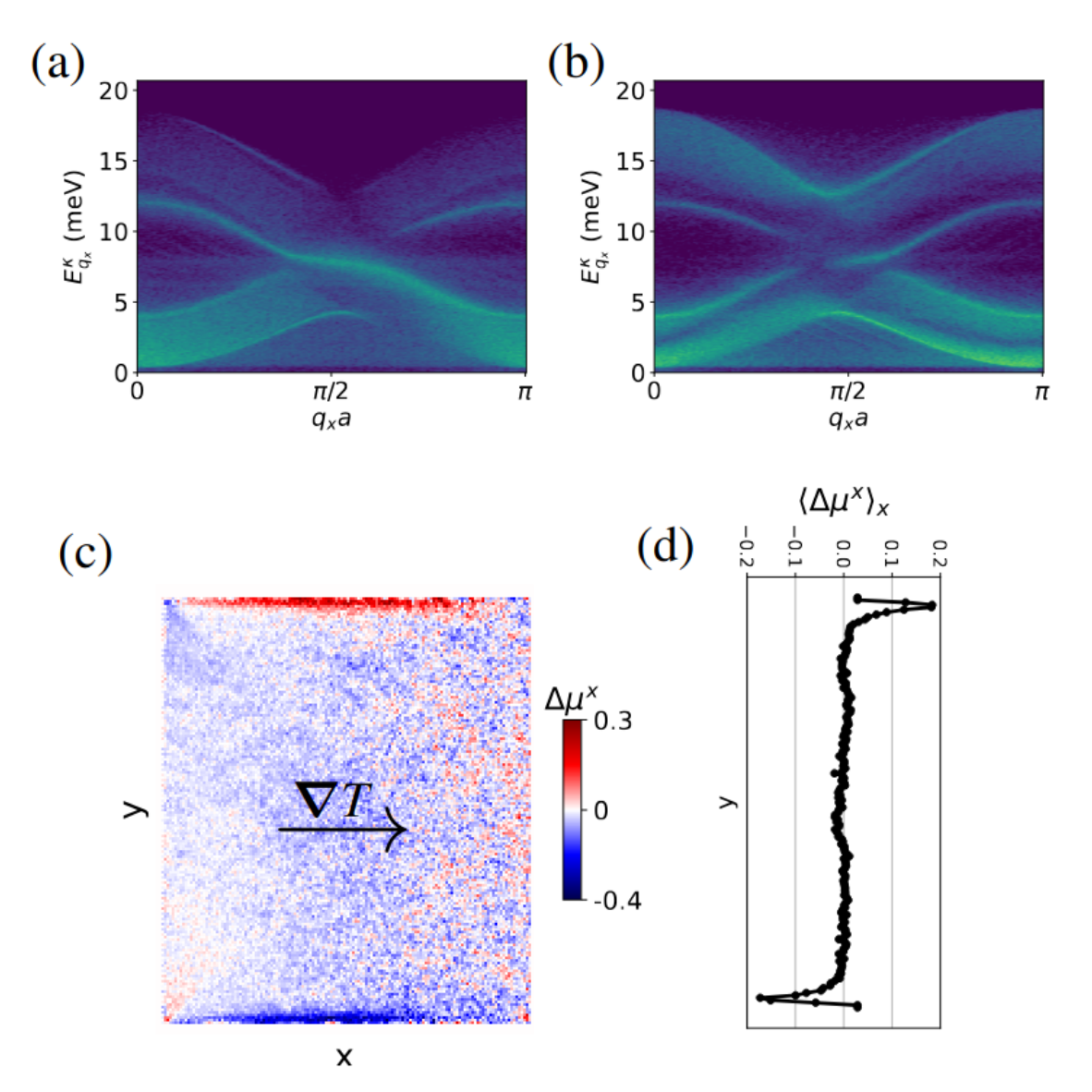}
    \caption{Response of the spin system to a temperature gradient along the $x$ direction in the Kitaev model with IP magnetization along the $x$ direction. (a) and (b) show the bulk magnon dispersion and the chiral edge modes of the upper zigzag edge (a) and lower zigzag edge (b) inside the topological magnon gap, which is shifted with respect to the $K$ symmetry point \cite{PRBcri3article}. (c) shows the 2D spatial distribution of the $x$ component of nonequilibrium spin accumulation. (d) shows the  spin accumulation as a function of the transverse $y$ direction, averaged along the longitudinal $x$ direction.}
    \label{fig:K_IP}
\end{figure}

As we showed in our previous study \cite{PRBcri3article}, in contrast to the DM spin model, tilting the magnetization IP along the $x$ direction in the Kitaev model does not close the magnon gap and the system remains topologically nontrivial. As presented in \cref{fig:K_IP}(a) and (b), the topological magnon gap in the bulk remains finite, though it is shifted in the magnetic Brillouin zone, for a detailed discussion see Ref. \cite{PRBcri3article}, and there are edge modes inside this topological magnon gap. We note that the chirality of the edge modes is reversed, compared to the OOP magnetization \cite{DMIandKitaevinCrI3analytical} which is a fingerprint of switching the sign of the Chern number and hence a topological phase transition.
In line with this swapped chirality of the edge modes, the nonequilibrium spin accumulation at the transverse edges is also the opposite, compare \cref{fig:K_IP}(c) and (d) to \cref{fig:K_OOP}(c) and (d). 
In order to transition from one topological phase, characterized by a Chern number, to the other, the topological gap must be closed and reopened. As reported in our previous study \cite{PRBcri3article}, although a magnon gap remains at the $K$ symmetry points for all magnetization directions, the position of the Dirac cones is shifted when we tune the magnetization direction from OOP to IP. The bulk topological magnon gap at the Dirac cones, which are no longer located at the $K$ symmetry points, closes at specific angles depending on the direction of tilting, and reopens as the tilt increases further, see Ref. \cite{DMIandKitaevinCrI3analytical} and SM in Ref. \cite{PRBcri3article}. This results in a topological phase transition, with its signature evident in the switching of the chiral edge modes. 
When the magnon gap is closed at certain tilting angles, the intrinsic spin Nernst effect goes to zero and there is no transverse spin signal.

\section{Concluding remarks} 
In this article, using atomistic spin simulations on CrI$_3$ as a prototype of 2D vdW magnetic materials, we investigated emerging chiral magnon edge states arising from DM and Kitaev spin models.
These edge states are protected by the effective time reversal symmetry, combination of the physical time-reversal symmetry and the inversion $C_2$ symmetry \cite{McClarty_2022}. 
Our results demonstrate that these edge states remain robust against nonlinear magnon interactions, which are inherently accounted for in our atomistic simulations. However, at higher magnon densities, where nonlinear interactions become stronger, the chiral edge states may break down \cite{MookBreakdownChiralEdgeModes}. 

We demonstrated two major differences in edge spin accumulation arising from the magnon spin Nernst effect for two generic spin models.
Utilizing a longitudinal temperature gradient, we may enhance the population of one chiral edge mode through the magnon spin Nernst effect. 

First, the nonequilibrium spin accumulation arising from the magnon Nernst effect is much higher in the DM model compared to the Kitaev model. We argue that this is due to the fact that our choice of the DM interaction respects the conservation of the spin angular momentum, while the Kitaev interaction does not and thus reduces the spin coherence length.
Second, there are clear qualitative differences. In the DM spin model, the magnon Nernst effect vanishes when the DM vector and the magnetization are orthogonal. In contrast, in the Kitaev model, the sign of the accumulation reverses when the magnetization is tilted from OOP to the IP $x$ direction. We expect an anisotropic IP response for the Kitaev model. Therefore, an angular measurement of the edge spin accumulation may be used as an indicator of the underlying microscopic characteristics of topological magnons.

Edge spin accumulation can be read out experimentally using optical detection techniques such as MOKE and XMCD \cite{Kato2004,expDetectionVDWmagnets}. The recently developed technique to measure magnons via nitrogen-vacancy centers in diamond \cite{NVcenters} is another  experimental method to measure edge states directly.
Furthermore, magnons in ferromagnets carry both spin angular momentum and heat. Thus, excluding other degrees of freedom such as phonons, the thermal Hall effect of magnons, which is easier to measure experimentally \cite{ExpThermHallCrI3}, could indirectly be a signature of the magnon spin Nernst effect and chiral edge states.

Our findings not only offer valuable insights into the topological and microscopic origins of spin interactions in 2D ferromagnetic vdW materials but also hold promising potential for applications in energy-harvesting Nernst devices \cite{mizuguchiEnergyHarvestingMatNernst} based on 2D materials and the next generation of spin qubits \cite{FukamiMAgnonQubitCoupling,ZhihaoMagnonsQuatumInformation}. Recently, a protocol was proposed for long-distance spin-qubit entanglement mediated by chiral magnons in 2D topological ferromagnets
\cite{MookLossQuBitcoupling}. We expect chiral magnons and topological magnons to play a significant role in advancing quantum information processing and enabling robust, long-distance qubit entanglement in future quantum technologies.

\textit{Acknowledgements --}
V.B. acknowledges A. Meo, P. Sukhachov, M. Weißenhofer, J. N. Kløgetvedt, E. J. G. Santos and R. F. L. Evans for helpful discussions.
This project has been supported by the Research Council of Norway through its Centres of Excellence funding scheme, Project No. 262633, ``QuSpin''. The Viking cluster was used during this project, which is a high-performance compute facility provided by the University of York. We are grateful for computational support from the University of York, IT services and the Research IT team. P.S. was supported by the Polish National Science Centre with Grant Miniatura No. 2019/03/X/ST3/01968

\bibliography{literature}

\onecolumngrid
\appendix 

\renewcommand{\thefigure}{S\arabic{figure}}
\renewcommand{\thetable}{S\arabic{table}}
\renewcommand{\thesection}{\arabic{section}}
\renewcommand{\thesubsection}{\Alph{subsection}}
\renewcommand{\theequation}{S\arabic{equation}}

\section{Additional numerical results} 

In the main text, we focus only on the zigzag edges of the system, that is, the upper and lower boundaries. In this SM, we present data for the armchair edges, i.e., left and right boundaries. Chiral edge states take a different path in the Brillouin zone on the armchair edges compared to zigzag edges \cite{MookLossQuBitcoupling}.
In order to study transport on these transverse edges, we apply a longitudinal temperature gradient along the $y$ direction. 

Furthermore, we present the topologically trivial DM model with IP magnetization. 

It is important to note that this study, without loss of generality, focuses exclusively on the tilting of magnetization within the $z-x$ plane. The topological gaps in the Kitaev model exhibit anisotropy and depend on the magnetization direction in the $x-y$ plane. A comprehensive analytical investigation into the angular dependence of the topological gaps can be found in Ref. \cite{DMIandKitaevinCrI3analytical}.
For our Kitaev model, the gap closes at \SI{35}{\degree} when the magnetization is tilted in the $z-x$ plane. However it is closed at \SI{90}{\degree} when tilting the magnetization in the $z-y$ plane.

\subsection{DM model with OOP magnetization: armchair edges}\label{sec:armchairEdges}
\begin{figure}
    \centering
    \includegraphics[trim = 10 30 10 10, clip,width=\linewidth]{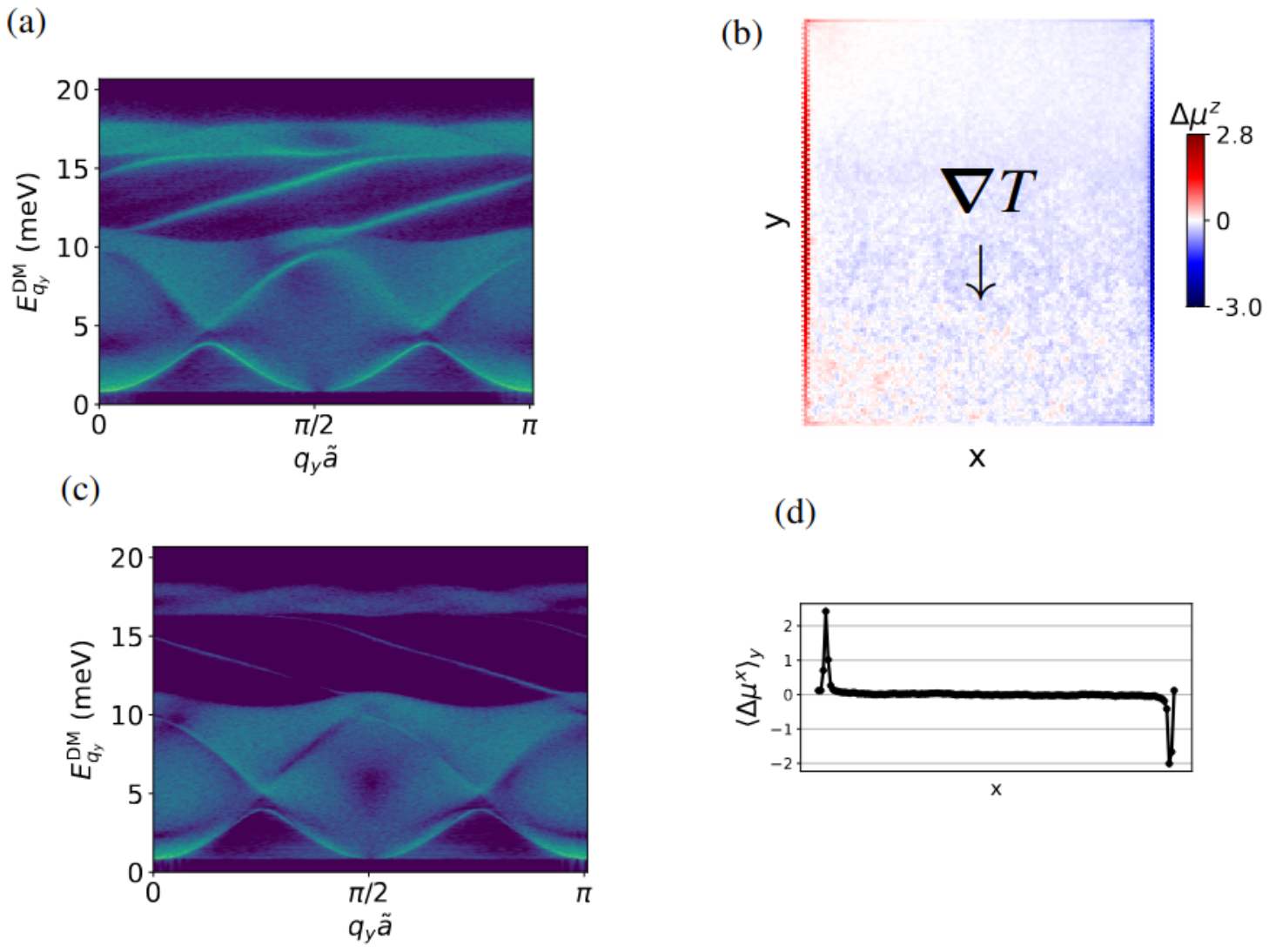}
    \caption{Magnon dispersion and nonequilibrium spin accumulation in the DM model with OOP magnetization. A longitudinal temperature gradient is applied along the $y$ direction to study the transverse armchair edges. 
    (a) and (c) show the bulk magnon dispersion and the chiral edge modes of the left armchair edge (a) and right armchair edge (c) inside the topological magnon gap. 
     The spatially resolved spin accumulation $\Delta \mu$ is shown in  (b) and (d).}
    \label{fig:DM_OOP_armchair}
\end{figure}

For the DM model with OOP magnetization, we can observe chiral edge modes inside the bulk gap in \cref{fig:DM_OOP_armchair}(a) and (b) and spin accumulation signal on the armchair edges in \cref{fig:DM_OOP_armchair}(c). Note that the Fourier transform is taken along the $y$ direction with step size $\tilde{a}=\sqrt{3}a$.

\subsection{Kitaev model with OOP and IP magnetization: armchair edges}

\begin{figure}
    \includegraphics[trim = 10 70 10 10, clip,width=\linewidth]{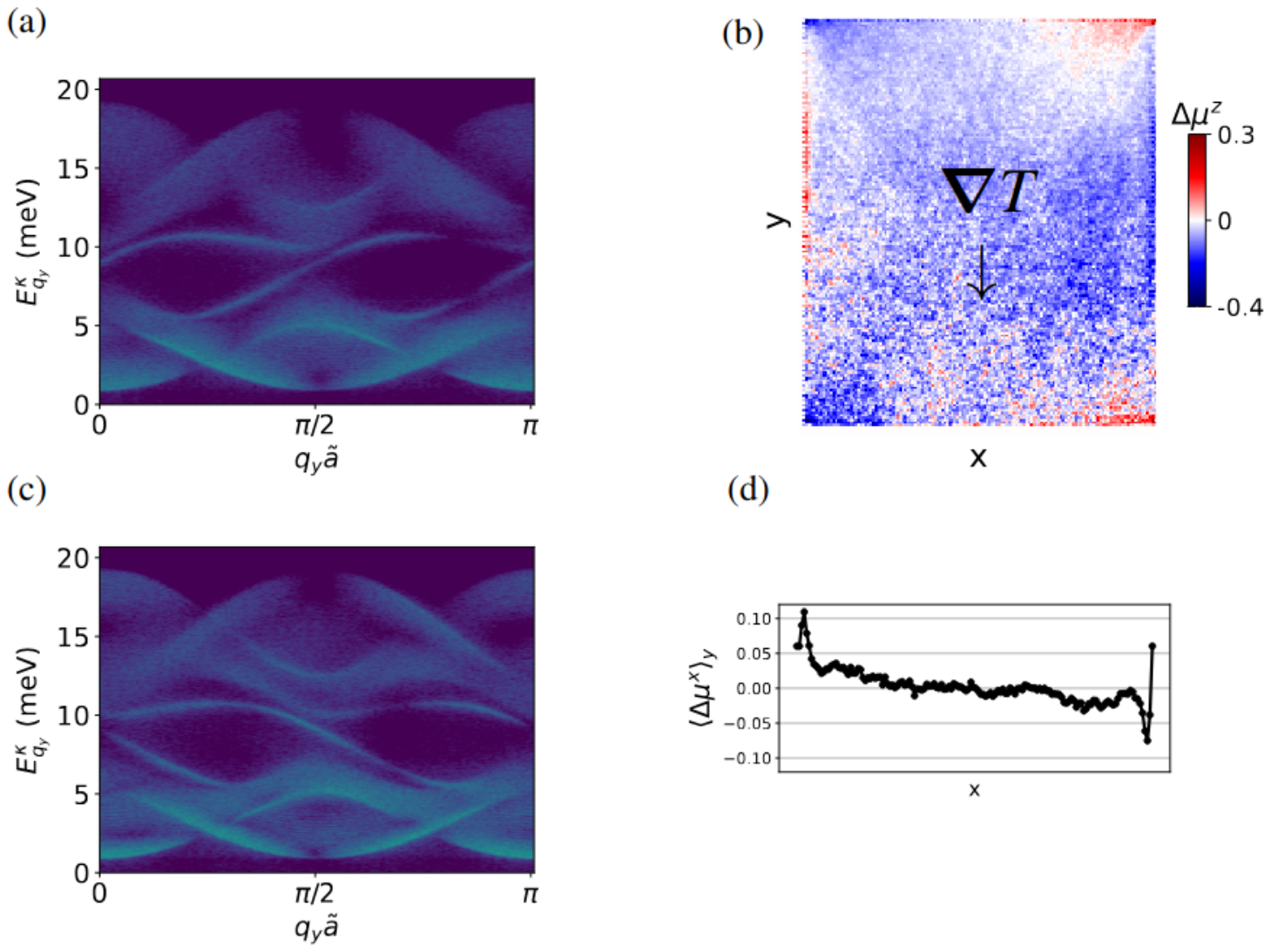}
    \caption{Magnon dispersion and spin accumulation in the Kitaev model with OOP magnetization. A longitudinal temperature gradient is applied along the $y$ direction to study the transverse armchair edges. 
    (a) and (c) show the bulk magnon dispersion and the chiral edge modes of the left armchair edge (a) and right armchair edge (c) inside the topological magnon gap. 
     The spatially resolved spin accumulation $\Delta \mu$ is shown in  (b) and (d).}
    \label{fig:Kitaev_OOP_armchair}
\end{figure}

\begin{figure}
    \includegraphics[trim = 10 70 10 10, clip, width=\linewidth]{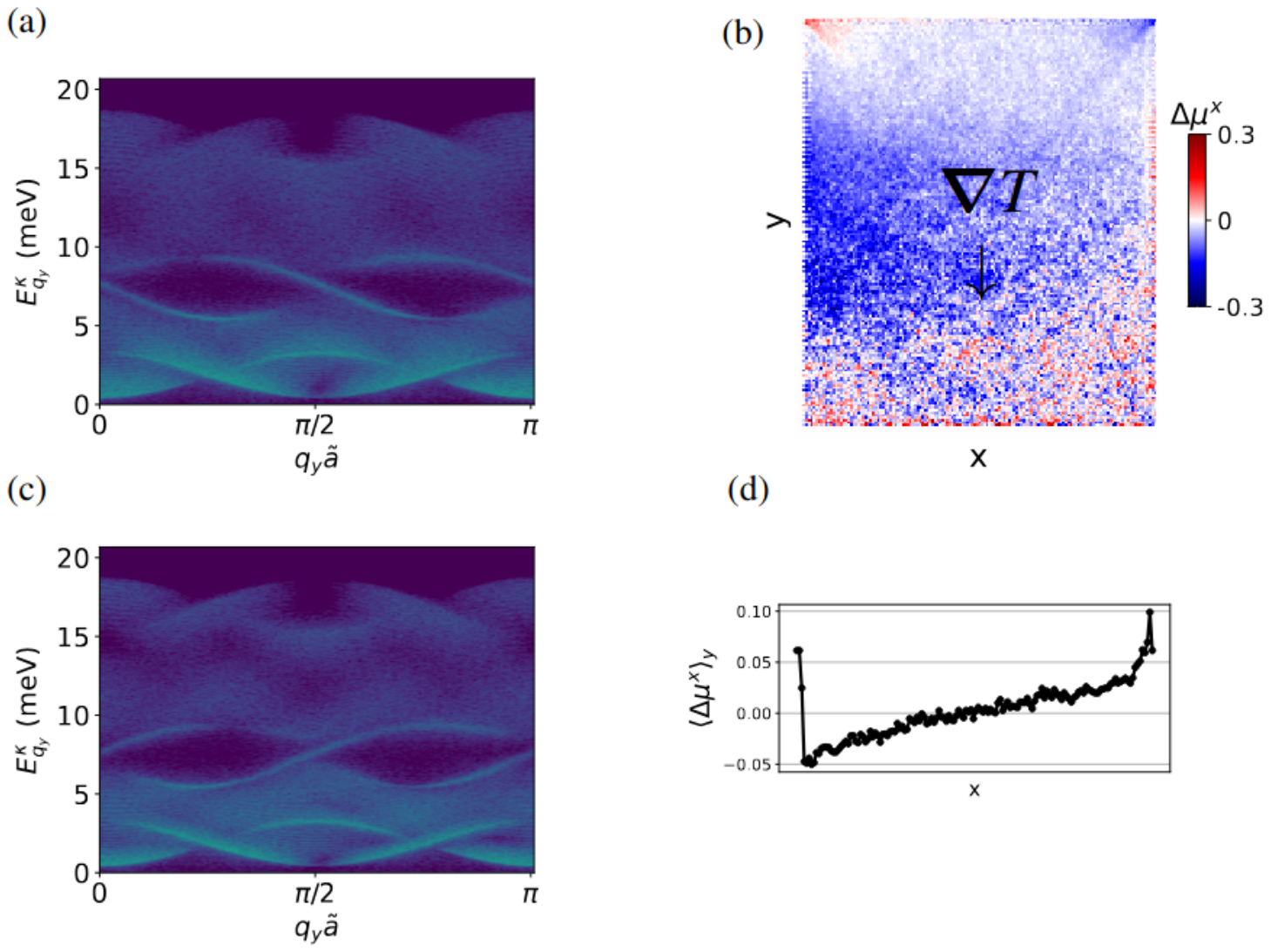}
    \caption{Magnon dispersion and spin accumulation in the Kitaev model with IP magnetization. A longitudinal  temperature gradient is applied along the $y$ direction to study the armchair edges. 
    (a) and (c) show the bulk magnon dispersion and the chiral edge modes of the left armchair edge (a) and right armchair edge (c) inside the topological magnon gap. 
     The spatially resolved spin accumulation $\Delta \mu$ is shown in  (b) and (d).}
    \label{fig:Kitaev_IP_armchair}
\end{figure}

As observed in the main text, the direction of the chiral edge modes changes in the Kitaev model when the magnetization is tilted from OOP to IP. This can also be seen on the armchair edges, compare \cref{fig:Kitaev_OOP_armchair}(a) and (b) to \cref{fig:Kitaev_IP_armchair}(a) and (b). This leads again to a switch of sign in spin accumulation at the chiral edges; compare \cref{fig:Kitaev_OOP_armchair}(c) to \cref{fig:Kitaev_IP_armchair}(c). Compared to the DM model, in the Kitaev model the armchair edges seem to suppress the spin accumulation even more compared to the zigzag edges and the signal is even weaker.

However, we observe an intriguing effect in the upper corners (low-temperature region) of the Kitaev system, both IP and OOP: the sign of the spin accumulation seems to switch between the zigzag and the armchair edge. Thus, there is a sharp contrast in the corner, which could indicate the emergence of corner states.

\subsection{DM model with IP magnetization: zigzag edges and armchair edges} \label{sec:DMwithIPmag} 
  
\begin{figure}
    \centering
    \includegraphics[width=\linewidth]{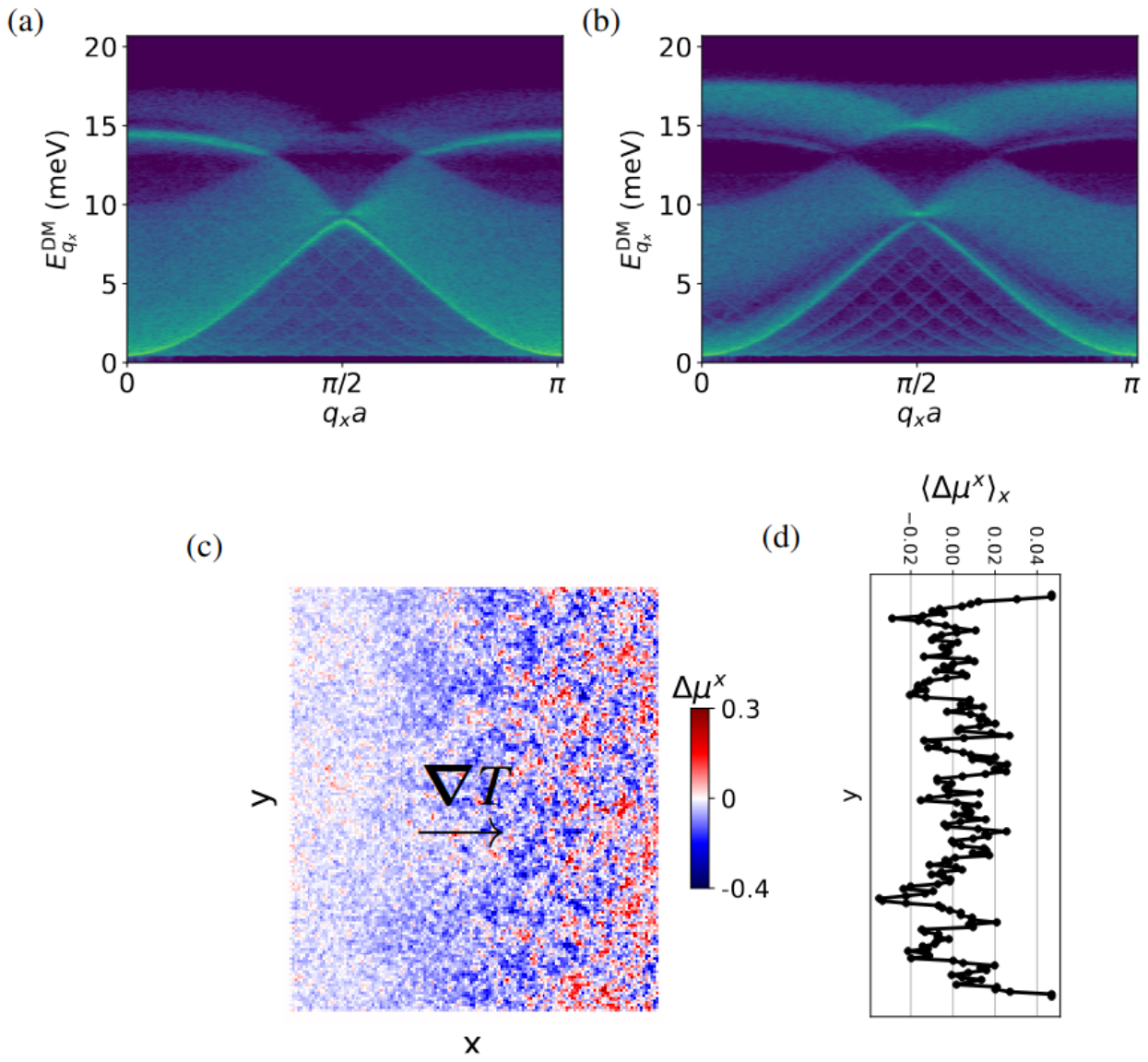}
    \caption{ Absence of edge spin accumulation in the DM model with IP magnetization. A temperature gradient is applied along the $x$ direction to study the transverse zigzag edges. The dispersion relations, left edge (a) and right edge (b), show that the bulk topological gap is closed. (c) and (d) show the absence of spin accumulation at the edges.}
    \label{fig:DM_IP}
\end{figure}

\begin{figure}
    \includegraphics[trim = 10 20 10 10, clip,width=\linewidth]{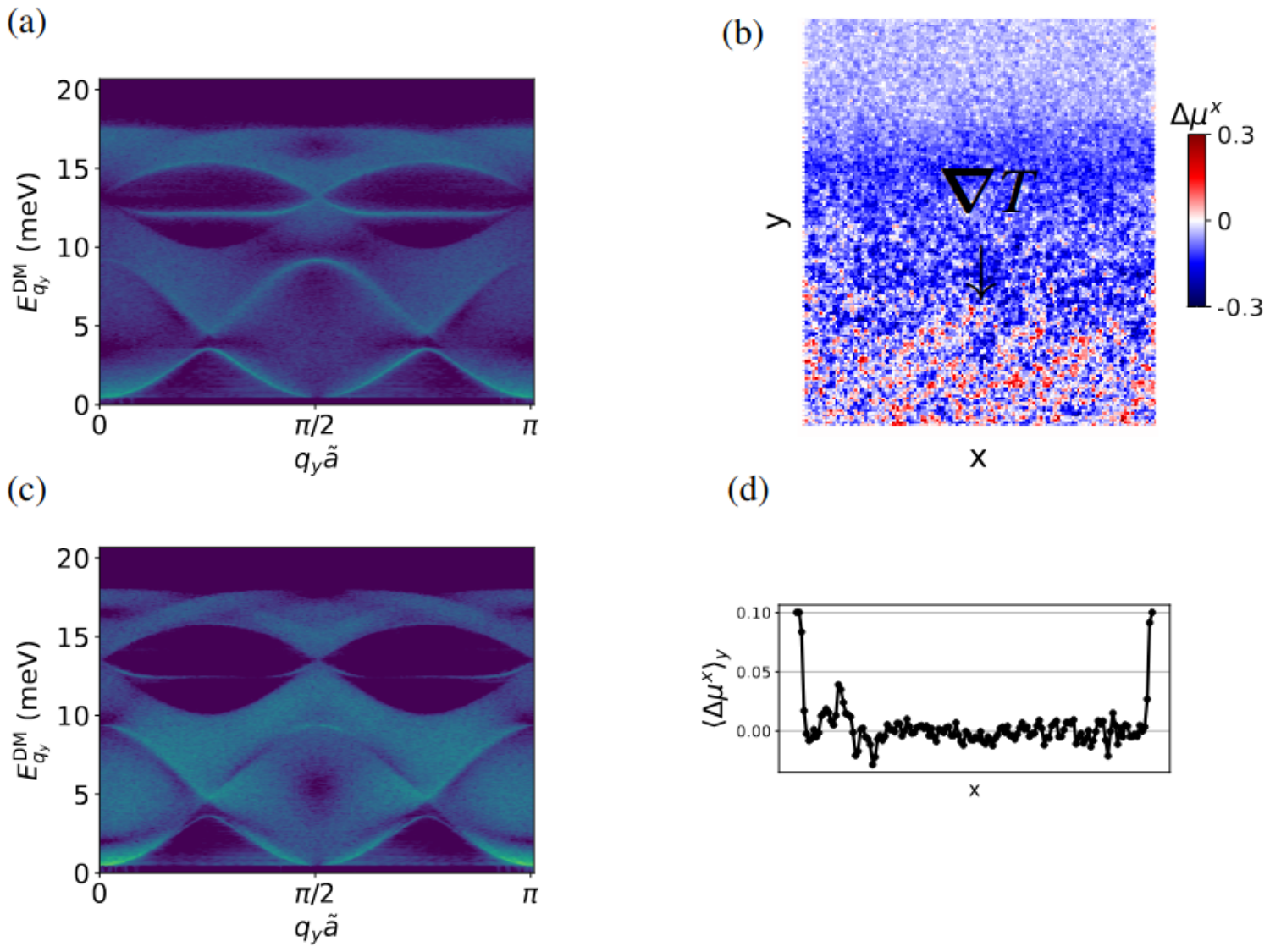}
    \caption{No magnon spin Nernst effect in the DM model with IP magnetization. A temperature gradient is applied along the y direction to study the armchair edges. The dispersion relations (a) and (c) show that the topological gap is closed. There is no transverse transport as shown with the difference in spin accumulation $\Delta \mu$ in top view (b) and in the transverse average (d).}
    \label{fig:DM_IP_armchair}
\end{figure}

When magnetization direction and DM vector are orthogonal, the magnon gap is closed in this model, and the system becomes topologically trivial. Thus, no chiral transport can be observed at the edges. This is indeed the case, as shown in \cref{fig:DM_IP} and \cref{fig:DM_IP_armchair}.

\section{Technical details } \label{sec:technicalDetails}
\subsection{Numerical methodology: atomistic spin simulations }
We use the atomistic spin simulation software package \vampire \cite{vampireURL,vampire} to simulate the magnetic moments on a hexagonal lattice of ferromagnetic CrI$_3$ layer. The stochastic Landau–Lifshitz-Gilbert (sLLG)
\begin{equation}
\label{eqn:LLG}
\frac{\partial \mathbf{S}_i}{\partial t} = -\frac{\gamma}{1 + \alpha^2} \left[\mathbf{S}_i \times \mathbf{B}^{\rm \kappa (DM)}_i + \alpha \mathbf{S}_i\times \left(\mathbf{S}_i \times  \mathbf{B}^{\rm \kappa (DM)}_i \right)\right],
\end{equation}
applied at the atomistic level~\cite{vampire,EllisLTP2015}, is solved numerically. Here, $\gamma$ is the electron gyromagnetic ratio and $\alpha$ is the Gilbert damping constant. 
The effective magnetic field for the Kitaev (DM) model $\mathbf{B}^{\rm \kappa (DM)}_i = - {\mu_s}^{-1}{\partial \mathcal{H}_{\rm \kappa (DM)}}/{\partial \sms_i} + \mathbf{\xi}^{\rm (th)}_i$, is composed of two terms: The first term is the deterministic contribution of the effective field, which is found from the corresponding spin-interaction Hamiltonian, and the second term is a stochastic thermal field, which introduces temperature to the system through an uncorrelated Gaussian thermal noise that obeys,
\begin{subequations}
\begin{align}
     &\langle\mathbf{\xi}^{\rm (th)}_i(t)\rangle =0,  \\ 
    & \langle\xi_{i,m}^{\rm (th)}(t) \xi_{j,n}^{\rm (th)}(t')\rangle = 2 \alpha k_\mathrm{B}T \gamma^{-1}\mu_s^{-1}\delta_{ij}\delta_{mn}\delta(t - t'), \label{eq:noise}
\end{align}
\end{subequations}
where $m,n = \{ x,y,z \}$ represent spatial components while $i,j$ denote the lattice points, and $k_\mathrm{B}$ is the Boltzmann constant.

\subsection{System setup and readout of the spin signal} 
The square system is $L_x=\SI{100}{nm} \times L_y=\SI{100.5}{nm}$ large, which corresponds to $N=46412$ sites. We consider a confined geometry with open boundary condition, which is armchair at the top and bottom edges ($x=0,x=L_x$) and zigzag along the right and left edges ($y=0,y=L_y$), see Fig.~1 of the main text. To compare the transport along these two types of edges, a temperature gradient $\bm{\nabla}T$ is applied along the $x$ direction and along the $y$ direction, separately. These directions we call the \textit{longitudinal} direction. If there is a chiral edge mode, a difference in spin signal will show up along the \textit{transverse} direction.

In order to increase the signal-to-noise ratio, we filter the transverse nonequilibrium signal, generated by the applied temperature gradient, from the thermal equilibrium background and from the longitudinal spin Seebeck signal. In order to to this, we run two sets of simulations per temperature gradient direction: one with positive slope, and one with negative slope. We then take the difference in spin accumulation in order to find $\Delta \mu$.

We consider a linear temperature gradient along the $x$ ($y$) direction where the local temperature $T(x)$ ($T(y)$) which is included in \cref{eq:noise} is given by 
\begin{equation*}T= \begin{cases}
    T_\mathrm{max}     & n_{x(y)} < L_{x(y)}/4 \\
    T_\mathrm{max} - c n_{x(y)}  & L_{x(y)}/4 \leq n_{x(y)} \leq 3L_{x(y)}/4 \\
    T_\mathrm{min}     & n_{x(y)} > 3L_{x(y)}/4
\end{cases}\end{equation*}
where $n_{x(y)}$ denotes the position along the $x$ ($y$) direction and $c$ the slope of the temperature gradient, which parametrize the change of temperature in units of Kelvin per lattice constant, K/$a$. 
To get a temperature gradient along the $-x$ ($-y$) direction, $T_\mathrm{max}$ and $T_\mathrm{min}$ must be exchanged and the sign of $c$ must be swapped. 

After letting the system reach the steady-state regime at about \SI{500}{ps}, the spin accumulation per site is recorded for \SI{5}{ns} with an output rate of \SI{1}{ps}.  $\dot{\bm{S}}_{i}$ which is needed for computing the spin pumping signal is calculated using the LLG time step which is \SI{1}{fs}. For each model and magnetization direction, an ensemble average with five stochastically independent realizations is conducted.

For the dispersion relations, the time-dependent spin configurations are recorded for \SI{500}{ps} with an output rate of \SI{0.1}{ps}. 

We set the Gilbert damping constant to $\alpha=0.001$.

\subsection{Temperature scaling}
\begin{figure}
    \centering
    \includegraphics[width=0.6\linewidth]{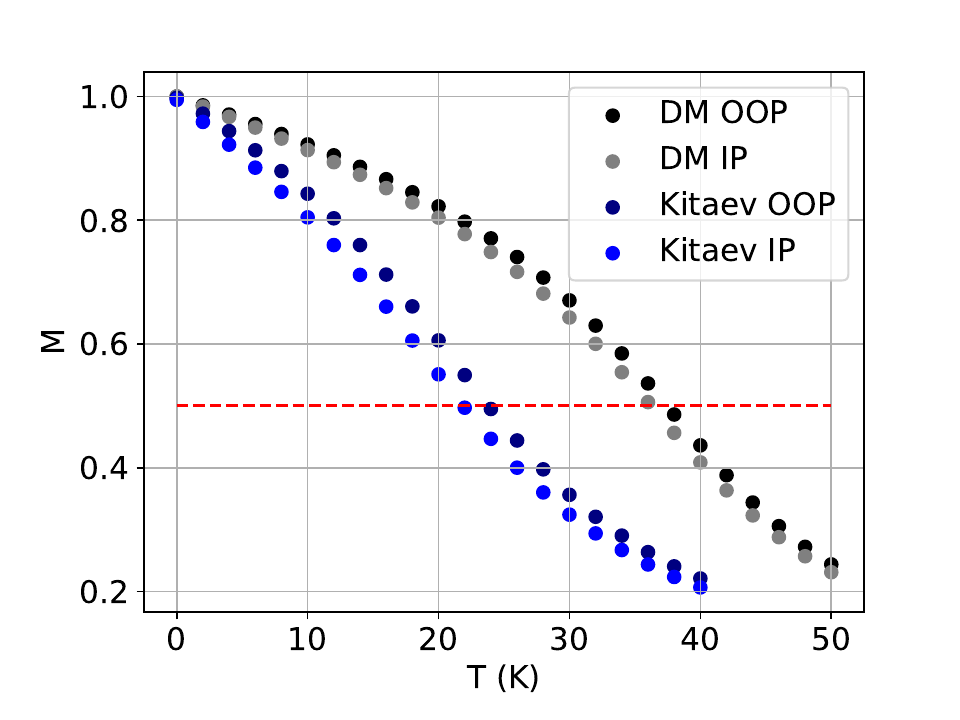}
    \caption{Temperature dependence of the normalized total magnetization $\mathrm{M}=\left\langle \sum_i^N S_i^{z (x)} \right\rangle_t/N$ for OOP and IP magnetic ground states and both spin models, computed with a Monte Carlo algorithm. The magnetization direction is set using a magnetic field of \SI{5}{T}. The reference temperature $T(\mathrm{M}=0.5)$ is used in simulations for normalizing the maximal and minimal applied temperatures in different setups.}
    \label{fig:Tcs}
\end{figure}

The Curie temperature of the system can vary based on the spin model and the direction of applied magnetic field, used to tilt the magnetization. We normalize the maximal and minimal temperatures for each simulation by $T(\mathrm{M}=0.5)$ so that the intensity of thermally excited magnons remains the same in all setups.
In \cref{fig:Tcs} the temperature-dependent magnetization is shown, and the temperature at which the magnetization reaches a value of $0.5$ is indicated with the red dashed line. We choose to set $T_\mathrm{max}/T(\mathrm{M}=0.5)=0.1$, leading to $T_\mathrm{max}^{\mathrm{DM,OOP}} = \SI{3.7}{K}$, $T_\mathrm{max}^{\mathrm{DM,IP}} = \SI{3.6}{K}$, $T_\mathrm{max}^{\kappa \mathrm{,OOP}} = \SI{2.4}{K}$ and $T_\mathrm{max}^{\kappa \mathrm{,IP}} = \SI{2.2}{K}$. Furthermore, we set $T_\mathrm{min}=0.01 T_\mathrm{max}$.

\subsection{Simulation parameters}
All spin parameters used in the atomistic spin simulations for the DM model, after Ref. \cite{Biquadratic_exchange_interactions_in_2D_magnets} are listed in \cref{tab:parametersDMImodel}, while the parameters for the Kitaev model, adapted from Ref. \cite{Hammel_FundamentalSpinInterac2020}, are provided in \cref{tab:parametersKitaevmodel}.
\begin{table}[h]
\begin{minipage}[t]{.45\linewidth}
      \caption{Atomistic spin simulation parameters for the DM model \label{tab:parametersDMImodel}}
      \centering
        \begin{ruledtabular}
        \begin{tabular}{c c c} 
        Parameter & Symbol & Value \\
        \hline 
         g-factor & $g_{e}$      & 2 \\ \hline
         Spin  & $S$ & $3/2$\\ \hline
         Magnetic moment & $\smmu=g_{e}S\muB$      & 3 $\muB$  \\ \hline
         Isotropic bilinear exchange & \makecell{ $J_1$ \\ $J_2$ \\ $J_3$ }  & \makecell{\SI{1.0}{meV} \\ \SI{0.32002}{meV} \\ \SI{0.0081}{meV}} \\ \hline
         Anisotropic bilinear exchange & \makecell{$\lambda_1$  \\ $\lambda_2$\\ $\lambda_3$}     & \makecell{ \SI{0.1068}{meV} \\\SI{-0.01024}{meV} \\ \SI{0.0091}{meV}} \\ \hline
         Biquadratic exchange & $K_{bq}$        & \SI{0.21}{meV} \\ \hline
        Easy axis anisotropy & $D_z$ & \SI{0.10882}{meV} \\  \hline
        DM strength & $A$ & \SI{0.31}{meV} \\
        \end{tabular}
        \end{ruledtabular}
    \end{minipage} \hspace{0.5cm}
    \begin{minipage}[t]{.45\linewidth}
      \centering
        \caption{Atomistic spin simulation parameters for the Kitaev model \label{tab:parametersKitaevmodel}}
        \begin{ruledtabular}
        \begin{tabular}{c c c} 
        Parameter & Symbol & Value \\
        \hline 
         g-factor & $g_{e}$      & 2 \\ \hline
         Spin  & $S$ & $3/2$\\ \hline
         Magnetic moment & $\smmu=g_{e}S\muB$      & 3 $\muB$  \\ \hline
         Isotropic bilinear exchange & $ J_0$   & \SI{0.55}{meV}  \\ \hline
        Easy axis anisotropy & $D_z$ & \SI{0.10882}{meV} \\  \hline
        Kitaev strength & $\kappa$ & \SI{4.5}{meV}
        \end{tabular}
        \end{ruledtabular}
        \end{minipage}
\end{table}

\end{document}